\documentclass[conference]{IEEEtran}

\usepackage{graphicx}
\usepackage{float}
\usepackage{makecell}
\usepackage{soul}
\usepackage{enumitem}
\usepackage{cite}
\usepackage{array}
\usepackage[hidelinks]{hyperref}
\usepackage{orcidlink}
\usepackage{balance}
\usepackage{xspace}
\usepackage{csquotes}

\hypersetup{
  pdflang=en,
  pdfauthor={Arno Leue, Akhila Bairy and Maike Schwammberger},
  pdftitle={In Terms of Explainability: Refining Requirements for Self-Explainable Systems},
  pdfkeywords={explanations, explainability, explainability engineering, explainability requirements, requirements engineering, explainable systems, taxonomy},
}

\makeatletter
\newcommand{\customlabel}[2]{%
  \protected@write \@auxout {}{\string\newlabel{#1}{{#2}{\thepage}{#2}{#1}{}}}%
  \hypertarget{#1}{#2}%
}
\makeatother

\begin{document}

\newcommand{\prop}[4]{%
  \vspace{3mm}
  \noindent
  \textbf{%
    Proposition 
    \customlabel{prop:#2}{#3}
    (#1):
  }
  \textit{#4}
}

\newcommand{\textbox}[1]{%
  \vspace{2mm}
  \noindent
  \fbox{%
    \hspace{.5mm}%
    \parbox{\dimexpr\columnwidth-2\fboxsep-2\fboxrule-1mm\relax}{%
      \vspace{0mm} #1 \vspace{0mm}
    }%
    \hspace{.5mm}%
  }
  \par
  \vspace{2mm}
}

\newcommand{\xreq}[4]{%
  \vspace{1mm}
  \noindent
  \setlist[description]{leftmargin=1\parindent,labelindent=0pt}
  \begin{description}
    \item[R\customlabel{xreq:#2}{#3}] #1: #4 
  \end{description}
  \vspace{1mm}
}

\newcommand{\xmp}[1]{\textit{#1}}

\newcommand{\scenario}[2]{%
  \item[\customlabel{scenario:#1}{S-#1}:] #2
}

\title{In Terms of Explainability: \\ Refining Requirements for Self-Explainable Systems}
\author{
  \IEEEauthorblockN{
    Arno Leue \orcidlink{0009-0009-6460-2928}, 
    Akhila Bairy \orcidlink{0000-0002-8796-1474},
    Maike Schwammberger \orcidlink{0000-0002-3344-6282}
  }
  \IEEEauthorblockA{
    \textit{Institute of Information Security and Dependability (KASTEL)} \\
    \textit{Karlsruhe Institute of Technology}\\
    Karlsruhe, Germany \\
    arno.leue@student.kit.edu, \\
    \{%
    \href{mailto:akhila.bairy@kit.edu}{akhila.bairy}, 
    \href{mailto:schwammberger@kit.edu}{schwammberger}%
    \}@kit.edu
  }
}
\maketitle

\begin{abstract}
Autonomous and software-intensive systems have been growing in occurrence, complexity, and assumed responsibility. 
Due to the high complexity of these systems, properties like transparency and explainability must be a focus of investigation. 
To date, no universally applicable definition and guide for the development of (self-)explainable systems exists. 
A need for explainability standards has already been recognized in the EU AI Act and the IEEE Transparency Standard 7001-2021.
To address this need, we propose unified definitions in explainability based on an analysis and combination of existing definitions. 
Additionally, we present structured explainability requirements that are necessary to build (self-)explainable systems.
By analysing the resulting taxonomy, we propose the incorporation of explanation goodness and thus correctness of explanations into the unified definitions. 
With our approach, we support the development of formal standards for \mbox{(self-)}explainable systems. Establishing such a uniform taxonomy also is a beneficial step towards certifying or auditing explainable systems.
\end{abstract}


\section{Introduction}\label{sec:intro}
Intelligent and semi-automated systems are deployed in more and more application domains. 
These application domains range from smart factories and smart home applications to (semi-)automated cars or automated delivery robots. 
At the same time, these software-intensive systems become larger, more complex and more opaque~\cite{Ber+23}, e.g. also due to an increase in Artificial Intelligence (AI) components.
Due to this lack of transparency~\cite{CKHS22}, it becomes even more difficult for stakeholders to understand aspects of system-behaviour~\cite{Blu+19}. 
Amongst other issues, a lack of understanding can contribute to unfairness or other negative impacts on the user~\cite{CBS21}.
Consequently, it is beneficial and essential for complex systems to explain their behaviour to ensure non-functional requirements, such as trust and understanding~\cite{Blu+19, Chaz23}.

The term \textit{explainability}~\cite{Koe+19} is closely connected to the terms interpretability, comprehensibility, understandability, or transparency~\cite{Chaz23, BDS23}. 
For clarity, we will stick to the differentiation of these terminologies made by Arrietta et al.~\cite{Arr+19}.
However, we generalise their focus on AI systems to more general types of software-intensive systems with the following notions:

\textit{Understandability} as the ability of a system to make a human understand (parts of) its behaviour and reasoning. 

\textit{Comprehensibility} as the ability of a system to present information in a human-understandable format.

\textit{Interpretability} as the ability of a system to provide meaning in understandable terms to a human.

\textit{Explainability}: as the ability of a system to provide information that helps a human understand a system.

Explainability being implemented into a system of interest, results in a self-explainable system.
Self-explainability itself is a crucial non-functional requirement that must be considered within system development processes~\cite{Chaz23}.
Such requirements describe perceived needs of stakeholders or capabilities a system should have, and thus form the basis of quality software.

In order to facilitate the engineering of the non-functional requirement of explainability, it is crucial to formally  define the term explainability. 
The first contribution of the paper at hand is an analysis of existing explainability definitions and a synthesis of unified explainability definitions from this investigation. 
Moreover, it is evident that explainability is not a simple, isolated, requirement; 
instead, additional requirements must be considered that enable or support the concept of explainability.
Our second contribution therefore is a structured analysis of requirements that support the non-functional requirement of explainability. 
We demonstrate our combined taxonomy and explainability requirements with a traffic management case study.

This paper is structured as follows; in Sect.~\ref{sec:sota}, we investigate relevant state in different aspects of explainability. 
In Sect.~\ref{sec:design}, we sketch our general approach. 
Our central contribution is twofold: 
a) We review explainability terminology and provide definitions for terms used in this paper in Sect.~\ref{sec:terminology}. 
b) We then derive and structure requirements that are necessary for integrating explainability into system design in Sect.~\ref{sec:requirements}. 
We demonstrate our findings from Sects.~\ref{sec:terminology} and \ref{sec:requirements} within a case study from the traffic management domain in Sect.~\ref{sec:case-study}. 
We close this paper in Sects.~\ref{sec:discussion}, \ref{sec:limits} and \ref{sec:conclusion} with a discussion, limitations, and a conclusion.

\section{State of the Art}\label{sec:sota}
We investigate the current state of the art on explainability definitions, aspects, quality, and standardisation efforts.

\paragraph{Definitions in Explainability}
The research field of explainability has brought forward different definitions of explanations and explainable systems~\cite{Ber+23, Chaz23, SMW24, Koe+19, CBS21}. 
Köhl et al.~\cite{Koe+19} explore explainability from the perspective of Requirements Engineering, based on which Chazette et al.~\cite{CBS21, Chaz23} propose a more detailed explainability definition.
Bersani et al.~\cite{Ber+23} propose increasing levels of explainability and a metric for quantitatively measuring a system’s explainability at a certain level. 
Based on this, Schwammberger et al.~\cite{SMW24} explore the integration of explainability levels into a framework for engineering run-time explainability.
The existing definitions will be compared in more detail in Sect. \ref{sec:terminology}.

\paragraph{Aspects of Explainability}
Sadeghi et al.~\cite{SKV21} propose explanation categories and a taxonomy of situations that they refer to as cases that need explanations. 
Additionally, they provide examples for these types of cases in order to be used to guide the requirements elicitation for the explanation capabilities of systems.
Furthermore, the proposed categories are envisioned as a means to extract explanation situations at the design time of a system.
Schwammberger and Klös~\cite{ScKl22} investigate explanation models and propose an extraction and refinement process for generating such models from system and environment models.
Bairy et al.~\cite{BHRS22} address the timing of explanations as a crucial aspect. 
They envision finding the optimal timing of an explanation, either before, during or after an explanandum, based on the attention level of the recipient.

\paragraph{Explanation Quality and Evaluation}
Brunotte et al.~\cite{BCKS22} observe that establishing standards for explainability allows for the creation of quality models for certification and auditing.
To that end, Chazette et al.~\cite{CKHS22} propose a framework that links dependencies, characteristics, and evaluation methods for explainability requirements. 
Their results show that it leads to the construction of explanations that increase usage, acceptance and satisfaction~\cite{CKHS22}.
Hoffman et al.~\cite{Hof+23} examine goodness of explanations and provide a checklist for it, based on which Schwammberger~\cite{Schw24} proposes a definition for explanation correctness and goodness.
The latter argues that trustworthiness of explanations can only be reached through a holistic Explainability Engineering process, which further complicates the achievement of goodness in explanations.

\paragraph{Explainability Taxonomies}\label{sec:related-work:taxonomy}
Similar to the quality framework by Chazette et al.~\cite{CKHS22}, proposals for taxonomies also exist in the field of explainability. 
Kahn et al.~\cite{Kah+02}, among others, propose a taxonomy for \textit{information}. 
Building on this, Nunes and Jannach~\cite{NuJa17} and Nauta et al.~\cite{Nau+23} propose taxonomies for \textit{explanations}. 
Several taxonomies for explainability have also been developed~\cite{ViLo21, Min+22, Lan+21-2}; however, mainly with a focus on explainable artificial intelligence (XAI). 
Speith~\cite{Spei22} examines the differences between taxonomies in XAI, and thus raises the need for standards again.

\paragraph{Explainability Standardisation}
Current standardisation efforts for intelligent and AI-supported software-intensive systems also include the need to integrate transparency and explainability mechanisms into such systems. 
This can, e.g., be found in the EU AI Act~\cite{EUAI24} and the IEEE standard 7001-2021 on transparency~\cite{IEEE7001}.
Apart from the need of integrating explainability into intelligent and software-intensive systems, a call for standards on \textit{how} to integrate explainability into systems can be observed in recent literature. 
As Brunotte et al.~\cite{BCKS22} conclude, one of the significant challenges related to explainability is the absence of a definitive and operational definition. 
Langer et al.~\cite{Lan+21} argue that the establishment of a parametrized definition or a meta-model of explainability could assist in defining aspects of explainability that are measurable and evaluable. 
Walke et al.~\cite{WBW23} conducted an analysis of the EU AI Act, leading to the conclusion that it provides only abstract regulations, thereby making it challenging to define specific metrics for achieving explainability. 

\section{Research Goal and Design}\label{sec:design}
We motivate and structure the key contribution of this paper in this section. 
A key finding from the research landscape is that explainability is a crucial non-functional requirement that must be integrated into software development processes~\cite{ChSc20, EUAI24, IEEE7001}. 
Furthermore, a wealth of different explainability definitions, aspects, and taxonomies has been introduced so far. 
However, the different taxonomies and definitions that we mention in Sect.~\ref{sec:sota}~d) have their merits.
  
The presence of numerous aspects of explainability in taxonomies and knowledge catalogues is of significant relevance. 
From a practical perspective, particularly from that of a software developer, functionalities such as explainability should adhere to established standards. 

The need for explainability also directly implies a need for some standards for integrating explainability into software-intensive systems. 
Such a standard should combine insights from different research directions. 
Thus we conclude the following two research needs RN\ref{rn:explainability} and RN\ref{rn:standards}.

\textbox{
  \setlist[description]{leftmargin=28pt,labelindent=0pt}
  \begin{description}
    \item[RN\customlabel{rn:explainability}{1}:]
    (Self-) Explainability of software-intensive systems
    
    \item[RN\customlabel{rn:standards}{2}:]
    Standardisation in engineering explainability
  \end{description}
}

To address the research needs RN\ref{rn:explainability} and RN\ref{rn:standards}, we venture towards a standardised and combined explainability terminology that conforms with existing explainability requirements. 
The following research questions RQ\ref{rq:definitions} -- RQ\ref{rq:benefits} guide our endeavour.

\textbox{
  \setlist[description]{leftmargin=28pt,labelindent=0pt}
  \begin{description}
    \item[RQ\customlabel{rq:definitions}{1}:] What are suitable definitions of explanations and explainability?
    \item[RQ\customlabel{rq:requirements}{2}:] What explainability requirements are needed for the implementation of (self-)explainable systems?
    \item[RQ\customlabel{rq:benefits}{3}:] How can insights from explainability requirements engineering (RQ\ref{rq:requirements}) be used to improve the proposed definitions?
  \end{description}
}

To answer our research questions, we first identify, combine and analyse existing terminologies in depth in Sect.~\ref{sec:terminology:constituents}.
Based on existing definitions of explanations and explainable systems we develop unified explainability definitions in Sects.~\ref{sec:terminology:expl_def}, \ref{sec:terminology:expl_sys_def}, \ref{sec:terminology:timing}. 
In Sect.~\ref{sec:requirements:elicitation}, we then identify, compare and analyse existing explainability requirements. 
Based on these, we generate a taxonomy in Sect.~\ref{sec:taxonomy} that emphasises interconnections between requirements for explanations based on definitions from existing explainability research and our own proposed definitions from Sect.~\ref{sec:terminology}. 
We investigate whether our proposed taxonomy meets the demands of explainability through an exemplary case-study in Sect.~\ref{sec:case-study}. 
We use the insights from Sects.~\ref{sec:terminology}, \ref{sec:requirements} and \ref{sec:case-study} to introduce a combined explainability definition in Sect.~\ref{sec:case-study:analysis}.

To illustrate concepts within the paper, we present the following example scenario that serves as a running example for the remainder of the paper: 
\xmp{
  Consider a smart coffee machine which is integrated into a smart home system. Residents can use the coffee machine via a smartphone application. 
  Users can query status information and communicate demands to the machine both through the application or manually at it.
}

\section{Terminology}\label{sec:terminology}
\begin{table*}[!t]
  \renewcommand{\arraystretch}{1.3}
  \caption{Comparison of existing definitions of Explanations.}
  \label{tab:explanations}
  \centering
  \setlength{\tabcolsep}{3pt}
  \begin{tabular}{ll}
    \hline
    \textbf{Source} & \textbf{Definition}: An explanation E for an explanandum X is a piece (of information I) that ... \\
    \hline
    Köhl et al.~\cite{Koe+19} & ... processed in context C makes any representative R of target group G understand X with respect to aspect Y. \\ 
    Bersani et al.~\cite{Ber+23} & ... makes X interpretable by a target group G. \\ 
    Chazette~\cite{Chaz23} & ... contributes to the addressee A’s understanding of X in context C. \\ 
    Schwammberger et al.~\cite{SMW24} & ... makes X understandable by a target group G with respect to an explainability goal $\theta$. \\ 
    \hline
  \end{tabular}
\end{table*}

In Sect.~\ref{sec:sota}, we provide an overview of existing explainability terminologies. 
In this section, we delve deeper into these terminologies and derive a unified definition from them.

\begin{figure}[h]
  \centering
  \includegraphics[height=2cm]{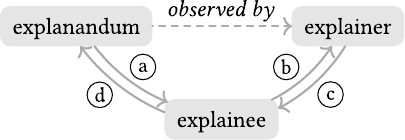}
  \caption{
    A minimal representation of the connections between constituents in the context of an explanation, which consists of an explainer, an explanandum, and an explainee.
    The explainer is the one who explains, the explanandum is what is being explained, and the explainee is the one to whom is explained.
    (a) is the matter of interest, 
    (b) the explainer realising this and therefore
    (c)~explaining it
    (d) so that the matter is understood better.
  }
  \label{fig:e-situation}
\end{figure}

In its simplest form, an explanation involves an \emph{explainer} providing information about a matter that is of interest to an \emph{explainee}, as visualised in Figure~\ref{fig:e-situation}.
The matter that is explained is usually referred to as the \emph{explanandum}.

Although this concept is only an abstraction of the process of explaining, it captures the essence of the key elements in an explanatory context. 
\xmp{
  Applied to the example, the explainer and the explanandum are the two systems: 
  usually, the explainer would be the smart home app, and the explanandum the smart coffee machine. 
  Depending on the situation, the relationship of the explainer and the explanandum can also be switched around.
}%
Next, we take a closer look at the constituents of an explanation to clarify the terminology used in this paper.

\subsection{Constituents of an Explanation}\label{sec:terminology:constituents}

According to IEEE standard 729-1983, a system S is a collection of components organised to accomplish specific functionality~\cite{ieee-se}. 
Systems can be classified according to various aspects: 
one such aspect is the type of components involved (e.g. software or physical components). 
The combination of both software and physical components constitutes the class of cyber-physical systems (CPS). 
In more detail, a CPS is a collection of independently interacting components that primarily transmutes how we interact with the physical world~\cite{cps-def}. 
Although systems in general are not limited to CPS, many explanation cases can be identified for CPSs. 
Example domains include robotics, autonomous driving and smart factories. 
\xmp{
  The smart coffee machine can be classified as a CPS. 
  This is due to its ability to autonomously and discretely decide on the software level, as well as to produce coffee in real-time with hardware using sensors and actuators. 
  The water tank filling with water and emptying (even at the same time) would be an example for a continuous CPS aspect.
}

An \textbf{explanandum X}, also known as a phenomenon~\cite{Wood79}, is a matter to be explained. 
This can be a system in general, or a specific aspect of a system (e.g. its reasoning processes, inner logic, intention, behaviour, decision procedures or knowledge)~\cite{CBS21}.
\xmp{
  Within our running example, an explanandum X could refer to the functionality of its internal decision-making algorithm, a specific aspect of coffee production, a self-cleaning routine, or a detail of its interaction with users.
}

The \textbf{need N} for an explanation of an explanandum X originates in a lack of understanding of it by an entity~\cite{Koe+19}. 
Thus, an \textbf{explainee}, also known as \textbf{addressee A}, is a recipient of explanation~\cite{Koe+19, Blu+19}. 
They are a subset of \textbf{stakeholders H}, where a stakeholder is anyone who has a vested interest in the entity.
All addressees are stakeholders, but not all stakeholders are addressees, since not all stakeholders are concerned with receiving an explanation.
Given that such a need N for an explanation is often expressed more than once, and potentially by several different individual explainees in a similar manner, these explainees are often classified into \textbf{target groups G}. 
Explanations that address a common need for an explanation are adapted to fit to a specific \textbf{representative R} of target group G that is equipped with the background knowledge and processing capabilities characteristic of the group~\cite{Koe+19}. 
\xmp{
  Specific target groups for the smart coffee machine could be consumers, maintenance workers, and software developers. 
  A specific representative of a group could be the owner of the coffee machine or an individual maintenance worker.
}

\textbf{Explainers} refer to a system or specific parts of a system that supply its explainees with the needed information~\cite{CBS21}. 
In more general terms, \textbf{means M} characterize the ability of an entity to provide and produce an explanation~\cite{Ber+23, Koe+19, SMW24}. 

In addition to explanandum, explainer, and explainee, a \textbf{context~C} and a goal of the explanation need to be considered. 
The context C is sometimes also referred to as the environmental entities~\cite{Ber+23} or 'environment' in general~\cite{CBS21}. 
In connection with systems as explainers, the context C of an explanation is set by entities from the environment that interact with the system, influence its behaviour and provide measurable factors that influence the observed explananda~\cite{SMW24}. 
Thus, the explainee and the explanandum are also included in the explanation context C.
A \textbf{goal $\theta$} of an explanation for a stakeholder group is a target to be reached by the process of explaining to the stakeholder~\cite{SMW24}.
The goal might also be included in the context.

\xmp{
  In our example, the entities that are comprised in the context C are primarily human stakeholders, environmental conditions and the dimension of time as additional factor. 
  Equally, the smart coffee machine system and the smart home system are also included in the explanation context. 
  Goals could relate to consumer satisfaction or a consumers trust in the system. 
  Economic and ecological optimisation, as well as minimising production delays, could also be (sub-)goals.
}

\subsection{Definition of Explanation}\label{sec:terminology:expl_def}

The literature that we analysed and compared in the previous section contains different definitions of explanations and explainability~\cite{Ber+23, Chaz23, SMW24, Koe+19, CBS21}.
Additionally, we also indicated that different phrases are used for similar concepts.
We briefly summarise and compare the core concepts of these different definitions of explanations in Table~\ref{tab:explanations}.

Different opinions exist among researchers w.r.t. the objectives that should be pursued with an explanation. 
As Miller~\cite{Mill19} already pointed out, researchers from philosophy, social and cognitive sciences have tried to give an answer to the nature of explanations for ages.

Some consensus seems to root in the fact that for most cases, there is not just one piece of information~\cite{Krn97,El-+19} that, once transmitted, clears the explanation need completely~\cite{Min+25}.
With this, \textit{explaining} is understood as a social process in which \textbf{information I} is transmitted multiple times in both ways, until the explanation need is sufficiently fulfilled~\cite{Roh+21}.
As such, explainability is not a technical concept but tightly coupled to human understanding~\cite{Koe+19}, and what makes a system explainable is the access to explanations~\cite{CBS21}. 

Based on this, we adhere to the phrasing of a single explanation as a \textit{contribution to understanding} as used by Chazette~\cite{Chaz23}. 
Note that this phrasing does not exclude the possibility that the goal of an explanation is to \emph{fully} explain something. 
We propose to keep the aspect of such a \textit{goal}~$\theta$ in our definitions, as suggested by Schwammberger et al.~\cite{SMW24}. 

\begin{figure}[H]
  \centering
  \includegraphics[width=.9\linewidth]{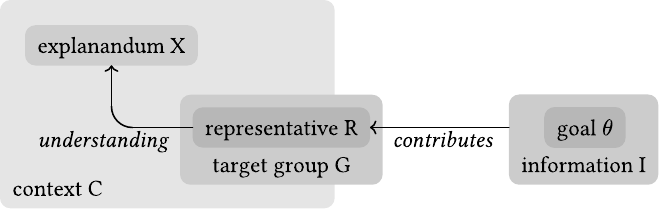}
  \caption{
    A visualisation for the proposed definition of an explanation E.
    The arrows are intended to be read as 'information' I contributes to the understanding of an 'explanandum' X by a 'representative' R.
  }
  \label{fig:proposed-e-def}
\end{figure}

This leads to a very central proposition in this paper: the definition of an \textbf{explanation E}. 
We derive the following proposition from the definitions of Köhl et al.~\cite{Koe+19}, Schwammberger et al.~\cite{SMW24} and Chazette~\cite{Chaz23} and visualise it in Figure~\ref{fig:proposed-e-def}.

\prop{Explanation}{explanation}{1}{
  Consider a representative R of a target group G. 
  An explanation E is a piece of information I that contributes to the representative R understanding an explanandum X in a context C with an explainability goal $\theta$.
}

\begin{table*}[!t]
  \renewcommand{\arraystretch}{1.3}
  \caption{Comparison of existing definitions of Explainable Systems.}
  \label{tab:explainable-systems}
  \centering
  \setlength{\tabcolsep}{5pt}
  \begin{tabular}{ll}
    \hline
    \textbf{Source} & \textbf{Definition}: S is explainable if and only if ... \\
    \hline
    \cite{Koe+19} & ... means M is able to produce an E in context C such that E is an explanation of X with respect to aspect Y, for target group G in context C. \\ 
    \cite{CBS21} & ... an entity E by giving information I enables an addressee A to understand an aspect X of S in context C. \\ 
    \cite{Ber+23} & ... it is able by a means M to produce an explanation E of an explanandum X for a target group G in context C. \\ 
    \cite{SMW24} &... it is able by a means M to produce an explanation E of an explanandum X for a target group G, with an explainability goal $\theta$ in context C. \\ 
    \hline
  \end{tabular}
\end{table*}

\subsection{Definition of Explainable System}\label{sec:terminology:expl_sys_def}

In the following, we enhance the definition of an \emph{explanation} E to that of an \textbf{explainable system $S_E$}. 
\textit{Explainability} refers to the ability to describe a system to other systems or humans, and thus can also be referred to as \textit{explainable systems}. 
Hence, we use \textit{explainability} and \textit{explainable systems} interchangeably. 
The concept of explainability can be defined in such a manner that its application is not exclusive to a particular entity, but that it is accessible to any entity. 
We focus on a definition of explainability that is limited to systems, which is in line with the definitions from~\cite{Koe+19, Ber+23, CBS21, SMW24}.
We briefly summarise the existing definitions in Table~\ref{tab:explainable-systems}.
  
In order to add the ability to explain to systems, the related works define a \textit{means M} as an asset that enables producing explanations~\cite{Koe+19, Ber+23, CBS21, SMW24}. 
Similarly, we adhere to the~phrasing.

The definition by Köhl et al.~\cite{Koe+19} differs from the other papers in that it specifies a particular \textit{aspect Y} of the explanandum that is addressed by an explanation. 
In the other papers, 'aspect' and 'explanandum' are used interchangeably and referred to as either term~\cite{Ber+23, SMW24, CBS21}.
Our definition aligns with the latter in that it encompasses the concept of an explanandum as either a system in general or a specific aspect of it.

Analogue to three of the four definitions~\cite{Koe+19, Ber+23, SMW24}, our approach also entails the utilisation of the definition of explanations within the definition of explainability. 
Our definition diverges from using the term 'information' by Chazette et al.~\cite{CBS21}, as we opt for the term 'explanation' to articulate the content being communicated. 
In addition, it is necessary to address the utilisation of an explainability goal, as previously defined in the context of explanations.

The following differentiation represents a change from all existing definitions: 
\textit{In the definition of explainability of systems, the explaining system and the explained system do not necessarily have to be the same}. 
In the following, we therefore refer to an explanation system $S_E$ and an \textbf{explanandum system $S_X$}. 
Since these two systems are each located in a context that does not necessarily have to be the same either, we also define the term of an \textbf{explanation context $C_E$} and an \textbf{explanandum context $C_X$}. 

\xmp{
  Using our example, we illustrate the difference between the explanation and explanandum context. 
  In the explanation context $C_E$, Bob tries to get a coffee from the smart coffee machine by entering his order at the designated terminal. 
  After several unsuccessful attempts, he gives up. 
  Later, he receives a notification on his smart home app informing him that the coffee order failed because the machine was still being cleaned.
  This situation provides the explanandum context $C_X$.
  The interactions with the explanation system $S_E$ and explanandum system $S_X$ took place in two different environments, at differed times and with different entities involved. 
  So the respective contexts are not the same.
}

\begin{figure}[H]
  \centering
  \includegraphics[width=.9\linewidth]{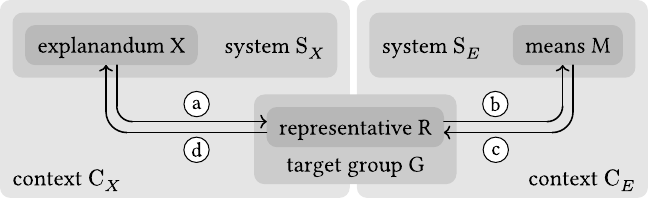}
  \caption{
    The proposed definition of explainable systems (XS), visualised. 
    (a) is the observation of the explanandum, 
    (b) refers to the expression of an explanation need. 
    (c) is the process of delivering an explanation, after which 
    (d), the understanding of the explanandum will be improved.
  }
  \label{fig:proposed-x-def}
\end{figure}

The approach of differentiating the two systems and contexts ensures that the definition of explainable systems is as general as possible. 
Additionally, it provides a suitable basis for defining \emph{self-explainable systems} and the \emph{timing of explanations}.
We phrase the definition of an explainable system in Proposition~\ref{prop:explainable-system} and provide a visualisation of it in Figure~\ref{fig:proposed-x-def}.

\prop{Explainable System}{explainable-system}{2}{
  A system $S_E$ in a context $C_E$ is able by a means M to produce an explanation E of an explanandum X of a system $S_X$ in a context $C_X$ for a target group G with an explainability goal $\theta$.
}

Based on Proposition~\ref{prop:explainable-system}, the definition of explainability now extends to a few special cases concerning the relation of the explanation system $S_E$ and the explanandum system $S_X$.
If both systems are identical, this is referred to as 'self-explainability' (Proposition~\ref{prop:self-explainable-system}).
Even though it is not necessarily the case, with a self-explainable system, the contexts $C_E$ and $C_X$ may also be identical.
One could refer to this as a 'live' self-explainable system.

\prop{Self-Explainable System}{self-explainable-system}{2.1}{
  A system $S_X$ is self-explainable iff $S_X$ is explainable and $S_E = S_X$.
}

\subsection{Timing of Explanations}\label{sec:terminology:timing}
\begin{figure}[H]
  \centering
  \includegraphics[width=.9\linewidth]{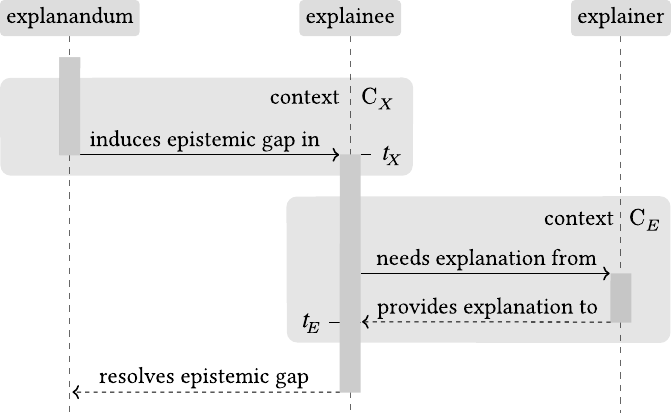}
  \caption{
    A sequential representation of a post-hoc explanation. 
    Time is represented along the vertical axis, increasing from top to bottom. 
    Both the explanandum context $C_X$ and the explanation context $C_E$ are depicted, as are the explanandum time $t_X$ and the explanation time $t_E$. 
    Due to $t_E > t_X$ in this example, it is a post-hoc explanation.
    The interactions correspond to arrows (a)–(d) as shown in Fig.~\ref{fig:proposed-x-def}.
  }
  \label{fig:proposed-x-seq}
\end{figure}

As we indicated in Sect.~\ref{sec:sota}, explanations can be provided before (ante-hoc), during or after (post-hoc) an event occurred. 
Consequently, the \textit{timing of an explanation} can be defined through the relationship between the explanation context $C_E$ and the explanandum context $C_X$. 
In case that the \textbf{timing $t_{E}$ of an explanation} precedes the \textbf{timing $t_{X}$ of its explanandum}, this relation describes an ante-hoc explanation. 
We express this through $t_{E} < t_{X}$. 
Analogously, an explanation occurring post-hoc ($>$) or during ($=$) an explanandum can be defined. 
Figure \ref{fig:proposed-x-seq} illustrates the time relation between $t_E$ and $t_X$. 
We provide the definition for this explanation timing in Proposition~\ref{prop:explanation-timing} and give an exemplary visualisation in \ref{fig:proposed-x-seq}.

\prop{Explanation Timing}{explanation-timing}{3}{
  The explanation E of an explanandum X of system S can be timed either before ($t_{E} < t_{X}$), during ($t_{E} = t_{X}$) or after ($t_{E} > t_{X}$) X happened. 
}

\section{Explainability Requirements}\label{sec:requirements}
In previous research, a wealth of requirements for explainability were identified with different levels of detail~\cite{CKHS22, NuJa17, Nau+23, Kah+02, Spe+24, SoFl20}.
Based on these requirements, Leue~\cite{Leue25} identified a subset of them regarding the explainability of software systems with a focus on the MAB-EX framework. 
It is evident that a substantial proportion of the requirements identified are of an optional nature and, consequently, are not universally necessary for self-explainable systems. 
Accordingly, we distilled nine general requirements through an analysis their results and discuss them in this section.

Requirement R0, \emph{explainability in systems}, is a high-level meta-requirement; all others are sub-requirements of this one.

\xreq{Explainability}{explainability}{0}{
  The explanandum system $S_X$ must be explainable for a target group G in a context C with respect to an explanandum X~\cite{Koe+19}.
}

\subsection{Requirements Elicitation}\label{sec:requirements:elicitation}

In the following, we elicit the requirements that emerged from distilling the requirements proposed by Leue~\cite{Leue25}.
They are presented individually at this stage, without any interpretative structure being imposed. 
Thus, their enumeration is merely for reference. 
Interdependencies between the requirements are discussed in the subsequent Sect.~\ref{sec:taxonomy}.

An important abstract requirement for explainability is the existence of a dedicated means for the purpose of producing information. 
Depending on the context, such means may be subsystems or entire systems.
This requirement can be referred to as explanation \textit{producibility} (R1).

\xreq{Explanation Producibility}{producibility}{1}{
  The explanation system $S_E$ must have means M to be able to produce an explanation~E~\cite{SMW24, Koe+19, Ber+23, CBS21}.
}

\noindent
Furthermore, the product must also contribute to the understanding of a target group in the explanandum in order to fulfil the goals of an explanation. 
This is described in explanation \textit{understandability} (R2).
We will examine what it means to understand an explanandum in more detail in Sect. \ref{sec:taxonomy}.

\xreq{Explanation Understandability}{understandability}{2}{
  The information I produced by the explanation system $S_E$ must contribute to the understanding of the explanandum X by any representative R of the target group G~\cite{SMW24, CBS21, Koe+19}.
}

\noindent
Two aspects of information that are already present in existing taxonomies of information and explanations are the type of \textit{presentation} (R3) (i.e., syntax) and the type of \textit{content} (R4) (i.e., semantics). 
Both should be adaptable by the explanation system to a stakeholder in order to contribute as much as possible to the understanding of the explanandum. 
Necessary attributes that implicitly play a role here are the \textit{comprehension} of presentation and the \textit{interpretation} of content. 
To ensure these, more detailed aspects of information play a role:

In terms of syntax, these are properties such as the type of \textit{modality} used~\cite{CKHS22, NuJa17}, the \textit{amount} of the types of modalities~\cite{CKHS22} and the \textit{general amount} of the information~\cite{Kah+02, Nau+23} as well as its \textit{coherence}~\cite{Nau+23} and \textit{consistency}~\cite{Kah+02, Nau+23}.

In terms of semantics, these include the \textit{relevance} of information for the explanation goal~\cite{Kah+02, Nau+23}, the \textit{contrastivity}~\cite{CKHS22, Nau+23}, \textit{complexity}~\cite{Kah+02, Nau+23}, and \textit{completeness}~\cite{Kah+02, Nau+23} of information, and the \textit{style}~\cite{CKHS22, NuJa17} of it. 
However, in order to remain at a meta-level here, we have not listed these individually as sub-requirements.

\xreq{Explanation Presentation}{presentation}{3}{
  The presentation (syntax) of the explanation E produced by the explanation system $S_E$ must be comprehensible by any representative R of the target group G in order to have any contribution to its understanding~\cite{Blu+19, Chaz23, CKHS22, NuJa17, Kah+02, Nau+23}.
}

\vspace{-2mm} 

\xreq{Explanation Content}{content}{4}{
  The content (semantics) of the explanation E produced by the explanation system $S_E$ must be interpretable by any representative R of the target group G in order to have any contribution to its understanding~\cite{CKHS22, NuJa17, Kah+02, Nau+23}.
}

\noindent
Another property of explanations that could also be considered part of the adaptability of content is correctness. 
Providing correct explanations for all explananda that need explanation is not possible due to models and idealisations.
In our case, we are dealing with explananda of systems that correspond to models from the ground up and can therefore be explained completely correctly by them. 
Explanation \textit{correctness} (R5) is thus an achievable characteristic. 
It is also desirable, since only correct information ultimately provides the basis for fulfilling explanation goals such as trust.

\xreq{Explanation Correctness}{correctness}{5}{
  The explanation E produced by the explanation system $S_E$ must be deducible from provably correct system models and context models~\cite{Schw24}.
}

\noindent
In addition to correctness, an explanation should also contribute measurably to the understanding of an explanandum as described above.
This enables monitoring whether an explanation goal has already been achieved or whether further information is needed.
Combined, these two characteristics can be defined as explanation \textit{goodness} (R6). 

\xreq{Explanation Goodness}{goodness}{6}{
  The explanation E produced by the explanation system $S_E$ must be correct and measurably help any representative R of target group G in understanding an explanandum X~\cite{Hof+23, Schw24}.
}

\begin{figure*}[!ht]
    \centering
    \includegraphics[width=.83\textwidth]{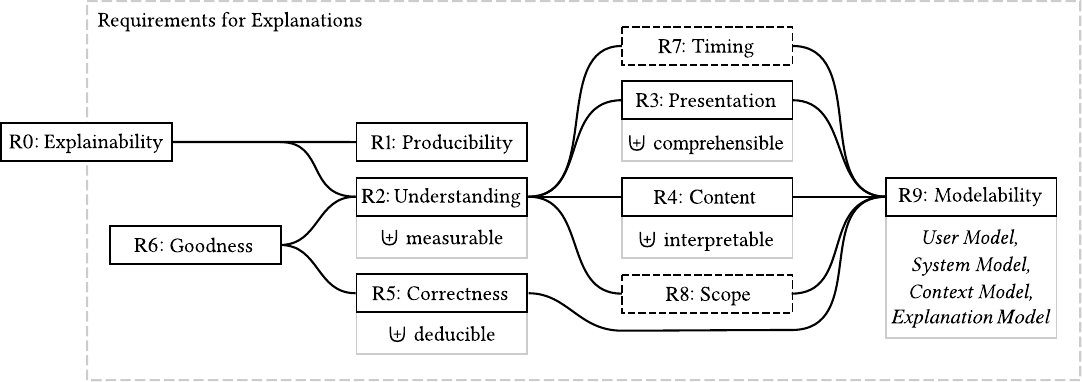}
    \caption{
        Taxonomy of abstract explanation requirements.
        Edges represent connections between the requirements themselves and also to the definition of explainability. 
        All requirements have 'Explanation' as a prefix. 
        The operator $\uplus$ is used as an indication that the following attribute is a property of the requirement it is mentioned with. 
        The connections are not hierarchical, though the graph can be read from left to right, e.g., "for 'Explanation Goodness', 'Explanation Correctness' is needed", based on the definition of goodness as mentioned in this chapter.
        The requirements 'Explanation Timing' and 'Explanation Scope' are dashed, because they could also be modelled as attributes or sub-requirements of 'Explanation Presentation' and 'Explanation Content' respectively.
    }
    \label{fig:xreq-graph}
\end{figure*}

\noindent
Two further dimensions that play a role in the provision of explanations are the choice of the right moment (timing, R7) and the choice of the information included (scope, R8). 
The former can also be considered a subcategory of the presentation of information, while the latter can be understood as a subcategory of content. 
However, both have a considerable influence on the understanding of stakeholders, hence we list them here.

\xreq{Explanation Timing}{timing}{7}{
  The explanation E produced by the explanation system $S_E$ should be adaptable in its timing, i.e. its provision before, during, or after explananda~\cite{BHRS22}.
}

\vspace{-2mm}

\xreq{Explanation Scope}{scope}{8}{
  The explanation E produced by the explanation system $S_E$ should be adaptable in its scope, i.e. providing local or global information on explananda~\cite{Ber+23}.
}

\noindent
Furthermore, a necessary aspect for the production of explanations is obtaining underlying information. 
In order to extract this, \textit{models} (R9) of relevant entities involved in an explanation are required, e.g. users, systems, context, or explanation history.
A system model can be an artefact from Software Engineering processes, e.g. architecture diagrams, communication protocols, etc, a context model describes the operating context of a system~\cite{ScKl22}.

\xreq{Explanation Modelability}{modelability}{9}{
  The representative R, context C, explanandum system $S_X$ and the explanation E itself should be modelable~\cite{SMW24, ScKl22, Schw24, Blu+19}.
}

\noindent
While provably correct models of systems are achievable, the complete correctness of context models poses a problem in terms of idealisations. 
Either way, correctness is highly dependent on the type of model and should therefore be chosen carefully.

\subsection{Requirements Taxonomy}\label{sec:taxonomy}
To find interconnections between the elicited requirements, we analysed conditions based on their wording or, if available, their definition itself. 
Starting with \textit{explainability} (R0), its definition calls for means for producing an explanation and thus for the \textit{producibility} (R1) of explanations. 
In addition, its transitive use of the definition of explanations requires the product (information) to contribute to the understanding of it, with which it implies \textit{understandability} (R2).

As already briefly touched on in Section \ref{sec:terminology}, there is no uniform definition of understanding.
However, as understanding is referred to as a process, we can make use of an abilities-based notion of understanding as proposed by Speith et al.~\cite{Spe+24},
which can be conditionalised as:

\vspace{3mm}
\noindent
\textbf{Condition \customlabel{cond:understanding}{1} (Understanding):}
\textit{
  A representative of a target group understands an explanandum iff they possess one or several understanding-related abilities.
}
\vspace{3mm}

With this, we can exploit at least two understanding-related abilities that are useful in understanding information: comprehensibility and interpretability, i.e. understanding the presentation and the content of information.
Additionally to their notion of understanding, Speith et al.~\cite{Spe+24} also proposed clusters of the understanding-related abilities, spanning from 'only' recognising content all the way to designing it.
From our point of view, the comprehension of information is located in the recognising cluster, and the interpretation of it somewhere in between the assessing and the design cluster.
Thus, as a beneficial differentiation, it makes sense to highlight that for \textit{understandability} (R2), it is necessary to be able to adapt the \textit{presentation} (R3) and \textit{content} (R4) of the explanation.   

There are more relevant aspects that are not directly connected to the understanding of an explanation, but also have an impact on how fast, good or complete it is understood. 
The \textit{timing} (R7) and \textit{scope} (R8) of an explanation are such that they are also interconnected with \textit{understandability} (R2). 

Another property of explanations that is of relevance is the correctness of the information that is used for an explanation, as only correct explanations are really wanted by stakeholders.
In combination, both explanation \textit{correctness} (R5) and \textit{understandability} (R2) are required for explanations to be classified as 'good'. 
This is defined in explanation \textit{goodness} (R6), which implies all three requirements to be connected as well.

As a final requirement, the ability of systems to build models to derive explanations from is needed~\cite{ScKl22, SMW24, Blu+19, Schw24}. 
This requirement, the explanation \textit{modelability} (R9), is interconnected to several other requirements, as different models are proposed to be of use for different steps in the explanation derivation process~\cite{ScKl22, Schw24}.
For the appropriate presentation of the syntax and semantics of content for a stakeholder, user, system, and context models are needed. 
Additionally, the correct timing of an explanation also implies the use of all these three models.
Explanation correctness, and with that goodness as well, also needs systems and context models in order to deduce correct information, as explicitly stated in its definition.

All of these interconnections are visualised in Figure~\ref{fig:xreq-graph}.
We explicitly do not want to highlight any hierarchy between them, although the connections can be interpreted as \textit{[left] needs [right]}.
\textit{Explainability} (R0) is intentionally placed at the border of requirements for explanations, as it can be seen as a high-level meta requirement. 
We decided to nevertheless place it into the figure, in order to highlight the direct connections to other requirements.
The set of requirements presented is meant to be a very high-level way of describing dimensions beneficial for explanations, and not as a complete set of requirements for the implementation of explainable systems.

\section{Case-Study}\label{sec:case-study}
We investigate whether our proposed taxonomy meets the demands of explainability through a case example in the following. 
For this purpose, we use the taxonomy of Sadeghi et al.~\cite{SKV21}, which structures explanation needs into different categories. 
We do not use the different explanation categories to point out additional ones, but merely to highlight different user requirements for explanations.
The setting of our exemplary case-study is a traffic management system at an urban intersection. 
The system controls traffic flow at the intersection and explains different situations and decisions to different involved stakeholders.

\subsection{Case Elicitation}

Overall, we investigate two different scenarios: in scenario \ref{scenario:1}, we consider an autonomous vehicle with a passenger, Alice, and in scenario \ref{scenario:2}, we consider an emergency vehicle, driven by Bob:

\setlist[description]{leftmargin=3\parindent,labelindent=\parindent}
\begin{description}
  \scenario{1}{
    On her way to work, Alice receives a notification in her autonomous vehicle. 
    The management system informs her that due to construction works on her usual route, it would be faster to take a different route. 
    Due to this, the car changes from the currently slower route to the new faster route. 
    The system decided to contact Alice before she were to notice the problem herself so that time would be saved and a cognitive overload for Alice would be avoided.
  }
  \vspace{2mm}
  \scenario{2}{
    On his way to the hospital, the emergency vehicle driver Bob approaches an intersection where the traffic situation is not clearly perceivable from a distance. 
    He contacts the management system to get a quick and reliable explanation of how to get through the junction as fast and safe as possible. 
    In order to save time, the system generates the information as a route rendered into the routing service of his vehicle. 
    Additionally, the system provides real-time images of the junction for Bob to be able to verify the planned route. This can help to increase Bob's ability to trust into the traffic management system.
  }
\end{description}

\subsection{Case Analysis}\label{sec:case-study:analysis}

The requirements identified for these two explanation cases can be classified into three categories: 
(a) technical requirements (e.g., the presence of an explainer and an explainee), which are not relevant for a broad approach to explainability, 
(b) abstract requirements, which describe generally relevant aspects, and 
(c) sub-requirements of those in b), that are not relevant at a high-level.
As such, in the following, we focus on requirements that were categorised in b).

Analysing the difference in \textit{presentation} (R3) and \textit{timing} (R7) of information provided in both cases, results in the requirement that both properties have to be dynamically adaptable to the stakeholder at runtime. 
This also implicitly highlights that information has to be \textit{producible} (R1).

It is evident that the \textit{content} (R4) of explanations should be adaptable in a way that it should be \textit{understandable} (R2) for the stakeholders involved. 

What is particularly emphasised in \ref{scenario:2} is that the explanation explicitly requested by the stakeholder must be based on correct information and should therefore also be correct itself. 
The main reason why an explanation is requested in \ref{scenario:2} is Bob's intent to cross the intersection safely. 
In order to enable and support trust in the system's decisions, it must be able to prove that the information and the explanations are correct. 
This confirms the need for the requirement for \textit{correctness} (R5) and, in combination with the required \textit{understandability} (R2), also the \textit{goodness} (R6) of explanations.

In \ref{scenario:1} and \ref{scenario:2}, it becomes evident that the explanation system (i.e. the traffic management system) has access to a model of the user to adapt the explanation in terms of presentation and content. 
This adaptation of the explanation can help to save mental space for the explainee. Additionally, the \textit{scope} (R8) of the explanation must be adaptable. 
Furthermore, it is suggested that the environment and the explanandum system, or at least its explananda, are captured within corresponding models. 
Consequently, the utilisation of a \textit{model} (R9) is necessary.

\section{Extending the term of Explanation}\label{sec:discussion}
Comparing the results of the case study with the taxonomy of explainability requirements proposed in Sect. \ref{sec:taxonomy}, all of the requirements mentioned there are at least confirmed, especially the correctness of information. 
Our definition of explainability in Sect. \ref{sec:terminology} is part of a basis for building connections between the requirements in the taxonomy.
Despite the existence of a need for explanation goodness, and thus also for explanation correctness, explanation goodness is not directly linked to the definition of explainability. 

However, we emphasise that this link should be present. 
We therefore integrate explanation correctness and the measurability of understanding, and thus overall goodness, into the definition of explanations.
In a transitive way, correctness and goodness are included in the definition of explainable systems.
The definition of explanations is therefore phrased as in Proposition~\ref{prop:new-explanation}.

\prop{Explanation}{new-explanation}{1.1}{
  Provably correct Information I that measurably contributes to a representative R of a target group G understanding an explanandum X in a context C with an explainability goal $\theta$.
}

\section{Limitations and Threats to Validity}\label{sec:limits}
Our work is exclusively based on qualitative data analysis. 
Consequently, there is a possibility that the results are affected by subjectivity during analysis. 
Thus, it is possible that some literature was not encountered when eliciting explainability and its requirements.
To eliminate this uncertainty, one could collect data on the basis of a systematic literature research. 
Next, we briefly discuss threats to the validity of the main parts of our research in more detail.

\paragraph{Terminology}
In order to compare existing definitions of explainability without the need for a detailed semantic analysis, we limited our selection to definitions that focus on formal terms.
The objective of this paper is not to provide a comprehensive comparison of \textit{all} notions of explainability; as such, this would have been outside the scope. 
Nonetheless, our structured requirement analysis serves as a frame for including explainability capabilities into system design processes.

\paragraph{Explainability Requirements}
In this paper, we propose an abstract set and taxonomy of requirements. 
However, we don't focus on the completeness of \textit{all} requirements for explanations. 
Instead, our main focus was on outlining the aspects in explainability, and identifying the structural connections between them.

\paragraph{Case Study}
Due to the small number of cases in the analysis of requirements from the stakeholder’s perspective in Sect.~\ref{sec:case-study}, not all requirements that could be used for self-explainable systems were analysed. 
However, a complete coverage of explanation requirements was not the goal in this research, but rather highlighting the importance of explanation correctness.

\section{Conclusion and Future Work}\label{sec:conclusion}
By comparing and analysing existing terminology, we developed a unified terminology for explanations (Prop. \ref{prop:explanation}) and explainable systems (Prop. \ref{prop:explainable-system}), which we propose as a starting point for developing standards in explainability.
In addition, we propose to integrate the requirements for explanation goodness into the definition of explanations (Prop. \ref{prop:new-explanation}), even though they are not necessary for explanations. 
Namely, we consider explanations based on false information to be undesirable, not least in terms of the goals of explainability.

By distilling requirements on explainable systems proposed by Leue~\cite{Leue25} and comparing them with existing taxonomies on information and explanations, we propose a taxonomy of requirements (Figure \ref{fig:xreq-graph}).
The requirements represent a minimal set that is needed for the implementation of (self-)explainable systems.
Additionally, the structure of the taxonomy highlights the logical interconnections between requirements, that could help as an abstract guideline for future research.

By analysing the structure of our requirements taxonomy (Figure \ref{fig:xreq-graph}), we highlight the need for explanation goodness to be incorporated into the definition of explanations.
This need was validated in a brief case-study (Sect. \ref{sec:case-study}), analysing explanation needs of stakeholders interacting with a management system.
In a similar way, more research on explainability requirements could be used to further refine definitions in explainability.

Based on the proposed definitions, future research should address further aspects of explanations in order to advance the development of explainability standards and requirements. 
To this end, a systematic literature review could be conducted to identify or include further aspects. 
Aspects from other disciplines should also be included, for example, from philosophy and the social sciences.

\section*{Acknowledgments}
\addcontentsline{toc}{section}{Acknowledgments}
The authors would like to thank the Ministry of Science, Research and Arts of the Federal State of Baden-W{\"u}rttemberg for the financial support of the project within the Innovation Campus Future Mobility.

\section*{Declaration of AI-Assisted Technologies}
During the preparation of this work, the authors used DeepL (https://deepl.com) for translation from German to English and for editing assistance to improve grammar and readability. 
The authors reviewed and revised all outputs from the tool and take full responsibility for the final content of the work.
\balance

\bibliographystyle{IEEEtran}

\end{document}